\documentclass[%
reprint,
superscriptaddress,
nobibnotes,
amsmath,amssymb,
aps,
floatfix
]{revtex4-2}

\usepackage[T1]{fontenc} 
\usepackage[]{graphicx}
\usepackage[utf8]{inputenc}
\usepackage[english]{babel}
\usepackage{blindtext}
\usepackage{lmodern}%
\usepackage{hyperref}%
\usepackage{xcolor}%
\usepackage{placeins}
\usepackage{bm}
\usepackage{tabularx}
\usepackage{booktabs}
\usepackage{notes2bib}
\usepackage[normalem]{ulem}
\usepackage{SIunits}

\newcommand{\mum}{$\mu$m}

\newcommand{\vg}{\ensuremath{V_{\rm G}}}
\newcommand{\sio}{SiO\ensuremath{_2}}

\newcommand{\dqmp}{Department of Quantum Matter Physics, University of Geneva, 24 Quai Ernest Ansermet, CH-1211 Geneva, Switzerland}
\newcommand{\gap}{Department of Applied Physics, University of Geneva, 24 Quai Ernest Ansermet, CH-1211 Geneva, Switzerland}
\newcommand{\unizurich}{Department of Chemistry, University of Zurich, CH-8057 Zurich, Switzerland}
\newcommand{\KW}{Research Center for Functional Materials, National Institute for Materials Science, 1-1 Namiki, Tsukuba 305-0044, Japan}
\newcommand{\TT}{International Center for Materials Nanoarchitectonics, National Institute for Materials Science, 1-1 Namiki, Tsukuba 305-0044, Japan}
\newcommand{\modena}{Dipartimento di Scienze Fisiche, Informatiche e Matematiche, University of Modena and Reggio Emilia, IT-41125 Modena, Italy}

\newcommand{\ie}{\emph{i.e.},}

\definecolor{AAcolor}{rgb}{0.7,0.1,0.4}

\usepackage{newfloat}

\definecolor{linkcol}{rgb}{0,0,0.4}
\definecolor{citecol}{rgb}{0.5,0,0}
\hypersetup{
	bookmarksopen=true,
	pdftitle="Quasi 1D electronic transport in a 2D magnetic semiconductor",
	pdfauthor="Wu et al",
	pdfsubject="Quasi 1D electronic transport in a 2D magnetic semiconductor", 
	pdfstartview={FitH}, 
	pdfmenubar=true, 
	pdfhighlight=/O, 
	colorlinks=true, 
	pdfpagemode=UseNone, 
	pdfpagelayout=SinglePage, 
	pdffitwindow=true, 
	linkcolor=linkcol, 
	citecolor=linkcol, 
	urlcolor=blue
}

\begin{document}

	\author{Fan Wu}\email{fan.wu@unige.ch}\affiliation{\dqmp} \affiliation{\gap}
	
	\author{Ignacio Gutiérrez-Lezama}\affiliation{\dqmp} \affiliation{\gap}
	
	\author{Sara A. Lopéz-Paz} \affiliation{\unizurich}
	
	\author{Marco Gibertini} \affiliation{\modena}
	
	\author{Kenji Watanabe} \affiliation{\KW}
	
	\author{Takashi Taniguchi} \affiliation{\TT}
	
	\author{Fabian O. von Rohr} \affiliation{\unizurich}
	
	\author{Nicolas Ubrig} \email{nicolas.ubrig@unige.ch}\affiliation{\dqmp} \affiliation{\gap}
	
	\author{Alberto F. Morpurgo} \email{alberto.morpurgo@unige.ch}\affiliation{\dqmp} \affiliation{\gap}

	
	\date{\today}
	
	
	\title{Quasi 1D electronic transport in a 2D magnetic semiconductor}
	
		\begin{abstract}
		
			We investigate electronic transport through exfoliated multilayers of CrSBr, a 2D semiconductor that is attracting attention because of its magnetic properties. We find an extremely pronounced anisotropy that manifests itself in qualitative and quantitative differences of all quantities measured along the in-plane \textit{a} and \textit{b} crystallographic directions. In particular, we observe a qualitatively different dependence of the conductivities $\sigma_a$ and $\sigma_b$ on temperature and gate voltage, accompanied by orders of magnitude differences in their values ($\sigma_b$/$\sigma_a \approx 3\cdot10^2-10^5$ at low temperature and large negative gate voltage). We also find a different behavior of the longitudinal magnetoresistance in the two directions, and the complete absence of the Hall effect in transverse resistance measurements. These observations appear not to be compatible with a description in terms of conventional band transport of a 2D doped semiconductor. The observed phenomenology --together with unambiguous signatures of a 1D van Hove singularity that we detect in energy resolved photocurrent measurements-- indicate that electronic transport through CrSBr multilayers is better interpreted by considering the system as formed by weakly and incoherently coupled 1D wires, than by conventional 2D band transport. We conclude that CrSBr is the first 2D semiconductor to show distinctly quasi 1D electronic transport properties.
		
	\end{abstract}

	\maketitle


\section{Introduction}

\begin{figure}
	\includegraphics[width=\linewidth]{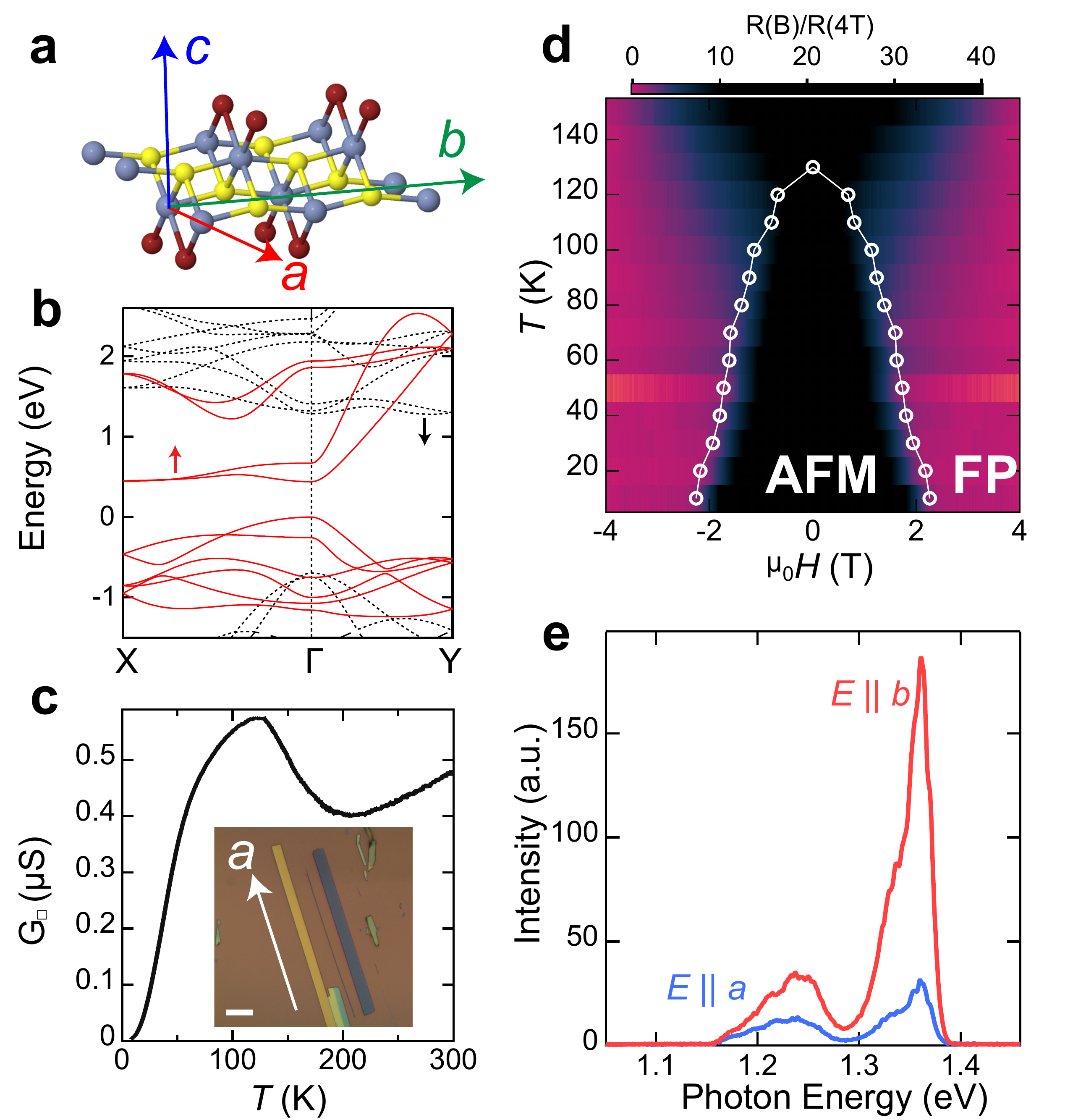}
	\centering
	\caption{\textbf{Known properties of CrSBr. } Using results of our work, we summarize what earlier studies have established about the electronic properties of CrSBr. \textbf{a)} Crystal structure of a CrSBr layer; the blue, yellow and red balls represent Cr, S and Br atoms. \textbf{b)}\textit{ ab-initio} calculation of the band structure of monolayer CrSBr, performed in the expected ferromagnetic ground state. The red solid and black dashed lines represent bands formed by spin up and spin down states, respectively. \textbf{c)} Temperature-dependence of the sheet conductance of a 6~nm-thick CrSBr device measured in the \textit{a}-direction. Inset: optical microscope image of a needle-shaped exfoliated CrSBr multilayer (the scale bar is 10 $\mu$m), showing that the crystalline orientation can be easily determined from its shape. \textbf{d)} Color plot of the resistance ratio $R(\mu_0 H)/R(\mu_0 H=4T)$ measured as a function of applied magnetic field and temperature. For each value of $T$, the white open circles mark the onset of positive magnetoresistance, and trace the boundary determined by the spin-flip transition, which separates the layered antiferromagnetic and the fully polarized phase. \textbf{e)} Polarization resolved photoluminescence spectra of a 6~nm thick, encapsulated CrSBr crystal (the red and blue lines represent the photoluminescence signal measured with polarization along the \textit{b}- and \textit{a}-direction).}
	\label{fig:01}
\end{figure}

Two-dimensional (2D) magnetic semiconductors are a unique platform to explore the interplay between magnetic and optoelectronic functionalities~\cite{burch_magnetism_2018,gong_two-dimensional_2019,gibertini_magnetic_2019,li_intrinsic_2019,mak_probing_2019,huang_emergent_2020}. The semiconducting properties of most 2D magnets investigated so far, however, are strongly affected by the extremely narrow widths of their conduction and valence bands, typically a few tens of meV or less~\cite{wang_electronic_2011,lado_origin_2017,tang_dopant_2017,mounet_two-dimensional_2018,li_spin-dependent_2019,menichetti_electronic_2019,zheng_ab_2019}. Such narrow bandwidths cause electron localization and prevent low-temperature conductivity measurements, which is why transport experiments probing the magnetic properties of 2D semiconductors have been so far limited to studies of tunneling through atomically thin multilayer barriers~\cite{klein_probing_2018,song_giant_2018,kim_one_2018,wang_very_2018,kim_tailored_2019,song_switching_2019,wang_determining_2019,long_persistence_2020}. CrSBr \cite{goser_magnetic_1990} (see \textbf{Figure~\ref{fig:01}a}) --a recently introduced 2D magnetic semiconductor-- appears to be an exception~\cite{telford_layered_2020,wilson_interlayer_2021}. First-principles calculations (shown in \textbf{Figure~\ref{fig:01}b}) predict its conduction band to have a width of approximately 1.5~eV~\cite{wilson_interlayer_2021,yang_triaxial_2021}. Accordingly, low-temperature in-plane magnetoresistance measurements (see \textbf{Figure~\ref{fig:01}c} and \textbf{d}) could be performed successfully, and analyzed to determine the magnetic phase diagram~\cite{telford_layered_2020}. The unique magnetic properties of this material have been further showcased by experiments on van der Waals (vdW) interfaces, in which CrSBr was found to imprint into an adjacent graphene layer a giant exchange interaction, much stronger than what has been reported in earlier work on analogous heterostructures~\cite{ghiasi_electrical_2021}. \\

With all the attention focused on its magnetic properties, the transport and optoelectronic response of CrSBr remains to be understood. Scanning tunneling spectroscopy measurements have shown that the Fermi level is very close to the bottom of --or possibly inside-- the conduction band and that charge carriers are electrons~\cite{telford_layered_2020}, but their microscopic origin (\ie\ what is the doping mechanism) and precise density are currently unknown. Rather uncommonly, the relatively large low-temperature conductivity --which confirms the presence of sizable non-intentional doping-- coexists with a bright fluorescence observed in photoluminescence experiments (see \textbf{Figure~\ref{fig:01}e})~\cite{wilson_interlayer_2021}. Additionally, the same band structure calculations (see \textbf{Figure~\ref{fig:01}b}) predicting a large bandwidth also show an extremely pronounced anisotropy: the bandwidth is about 1.5~eV in the $\Gamma-Y$ (which corresponds to the crystallographic \textit{b}-axis), but nearly vanishes in the perpendicular $\Gamma-X$ direction (\ie\ in the \textit{a}-direction)~\cite{wilson_interlayer_2021,yang_triaxial_2021}. Even though this prediction is somewhat surprising, because no obvious pronounced anisotropy can be identified in the atomic structure of the material (see \textbf{Figure~\ref{fig:01}a}), indications of anisotropy (see \textbf{Figure~\ref{fig:01}e}) were indeed found when measuring the photoluminescence polarization dependence~\cite{wilson_interlayer_2021}. However, it is not known whether low-frequency transport is also anisotropic, because previous studies only investigated the transport properties along the crystallographic \textit{a}-direction~\cite{telford_layered_2020,telford_hidden_2021}. Clearly, investigating more in depth the nature of transport and of the semiconducting properties of CrSBr is essential to understand their interplay with magnetism.\\

Here, we probe the direction-dependent transport properties of thin exfoliated CrSBr multilayers and the energy dependence of the density of states, and reveal an experimental phenomenology that cannot be reconciled with conventional 2D band transport of a doped semiconductor, \ie\ the scenario invoked to interpret past experiments. A key finding is that the temperature and gate voltage dependence of the conductivity measured along the \textit{a}- and \textit{b}-direction exhibit striking qualitative differences. In the \textit{a}-direction, the conductivity $\sigma_a$ decreases by two orders of magnitude upon lowering $T$ from 100 to 10~K, and is strongly suppressed (in some devices virtually fully suppressed) upon applying a sufficiently large negative gate voltage. In contrast, in the \textit{b}-direction the conductivity $\sigma_b$ is approximately the same at room temperature and at $T = 10$~K, and remains unaffected by the application of a gate voltage. The difference in transport along the \textit{a}- and \textit{b}-direction is so pronounced large that at $T = 10$~K and in the presence of negative gate voltage, the ratio $\sigma_b/\sigma_a$ ranges between $3\cdot10^2$ and $10^5$ depending on the devices considered in the analysis. Magnetotransport is also unusual: the sign of the low-$T$ magnetoresistance is opposite in the \textit{a}- and \textit{b}- directions, and the transverse magnetoresistance exhibits no sign of Hall effect irrespective of current direction and of temperature at which the measurements are performed (\ie\ above or below the Néel temperature). Finally, spectrally resolved photocurrent measurements reveal an abrupt increase of photocurrent above the light absorption threshold $E=E^*$ followed by a decrease proportional to $1/\sqrt{E-E^*}$, a characteristic manifestation of the \emph{van Hove} singularity of a one-dimensional (1D) band. All these observations consistently indicate that the electronic properties of CrSBr can be better understood by viewing the material as formed by incoherently coupled 1D chains, than by considering it as an (even strongly) anisotropic 2D electron system.\\

\section{Results and Discussion}
Our experiments rely on devices nano-fabricated using crystalline CrSBr multilayers exfoliated onto a doped Silicon substrate (acting as gate electrode), covered by a 285 nm thick \sio\ layer (a total of 14 devices have been fabricated and used in different measurements). Although CrSBr is considerably more air-stable than some of the magnetic materials investigated earlier~\cite{telford_layered_2020,wilson_interlayer_2021,ghiasi_electrical_2021}, stability is not perfect and prolonged air exposure over a few weeks still affects the multilayers strongly. Therefore exfoliation is performed in the controlled atmosphere of a glove-box and devices are either encapsulated in hexagonal boron nitride (hBN) or exposed to air for the shortest possible periods of time during the fabrication process (taking these precautions we have not seen clear differences between encapsulated and non encapsulated devices). The thickness of the multilayers --between 6 and 12~nm for most devices studied-- was determined by atomic force microscopy (AFM), and the crystallographic orientation was identified by selecting needle-shaped flakes whose long direction corresponds (as established in earlier work~\cite{lee_magnetic_2021,ghiasi_electrical_2021,wilson_interlayer_2021}) to the \textit{a}-direction (\textbf{Figure~\ref{fig:01}d}). Contacts to the multilayers were then attached in a Hall-bar configuration by means of electron-beam lithography, electron-beam evaporation of a Pt/Au bilayer and lift-off (for encapsulated devices, the hBN was reactively ion etched prior to evaporation of the metal; encapsulation was performed using by-now common pick-up and transfer techniques~\cite{wang_one-dimensional_2013}).\\

\begin{figure*}
	\centering
	\includegraphics[width=0.9\linewidth]{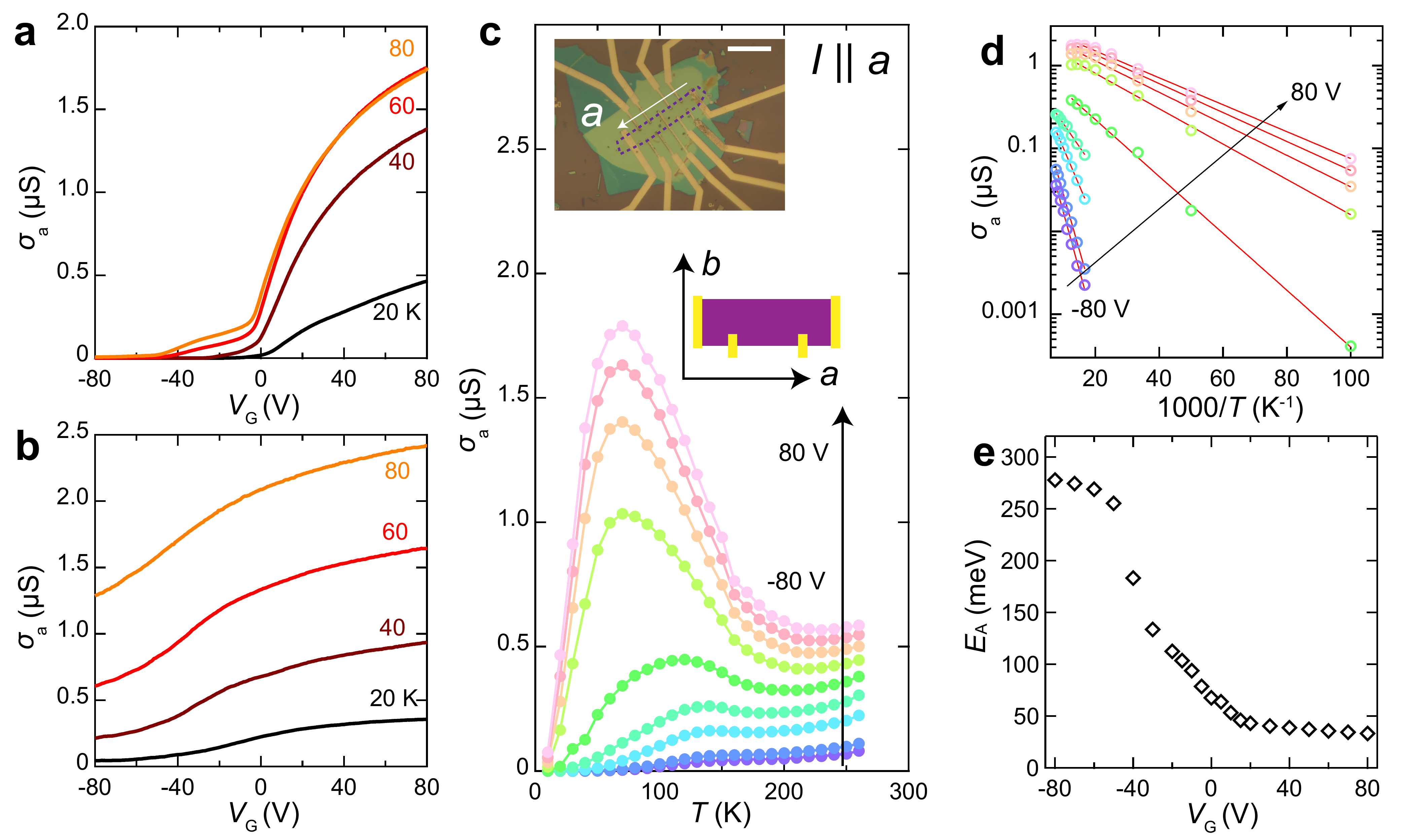}
	\caption{\textbf{Transport properties of CrSBr in the \textit{a} direction.} \textbf{a,b} Conductivity $\sigma_a$ of two different devices realized respectively on \textbf{a}) 6nm and \textbf{b}) 12 nm thick CrSBr multilayers, as a function of gate voltage, for different temperatures, $T $ = 20, 40, 60 and 80 K. \textbf{c)} Conductivity $\sigma_a$ versus temperature for the device whose data are shown in panel \textbf{a}, for different values of \vg\ (from -80 V to 80 V in 20 V steps). The insets show an optical microscope image of the device (the scale bar is 20 $\mu$m) and a schematics of the measurement configuration. \textbf{d)} Arrhenius plot of $\sigma_a$ as a function of inverse temperature for the same values of \vg\ shown in panel \textbf{c}. \textbf{e)} Activation energy extracted from the analysis of $\sigma_a$ as a function of gate voltage.}
	\label{fig:02}
\end{figure*}

\begin{figure*}
	\centering
	\includegraphics[width=0.9\linewidth]{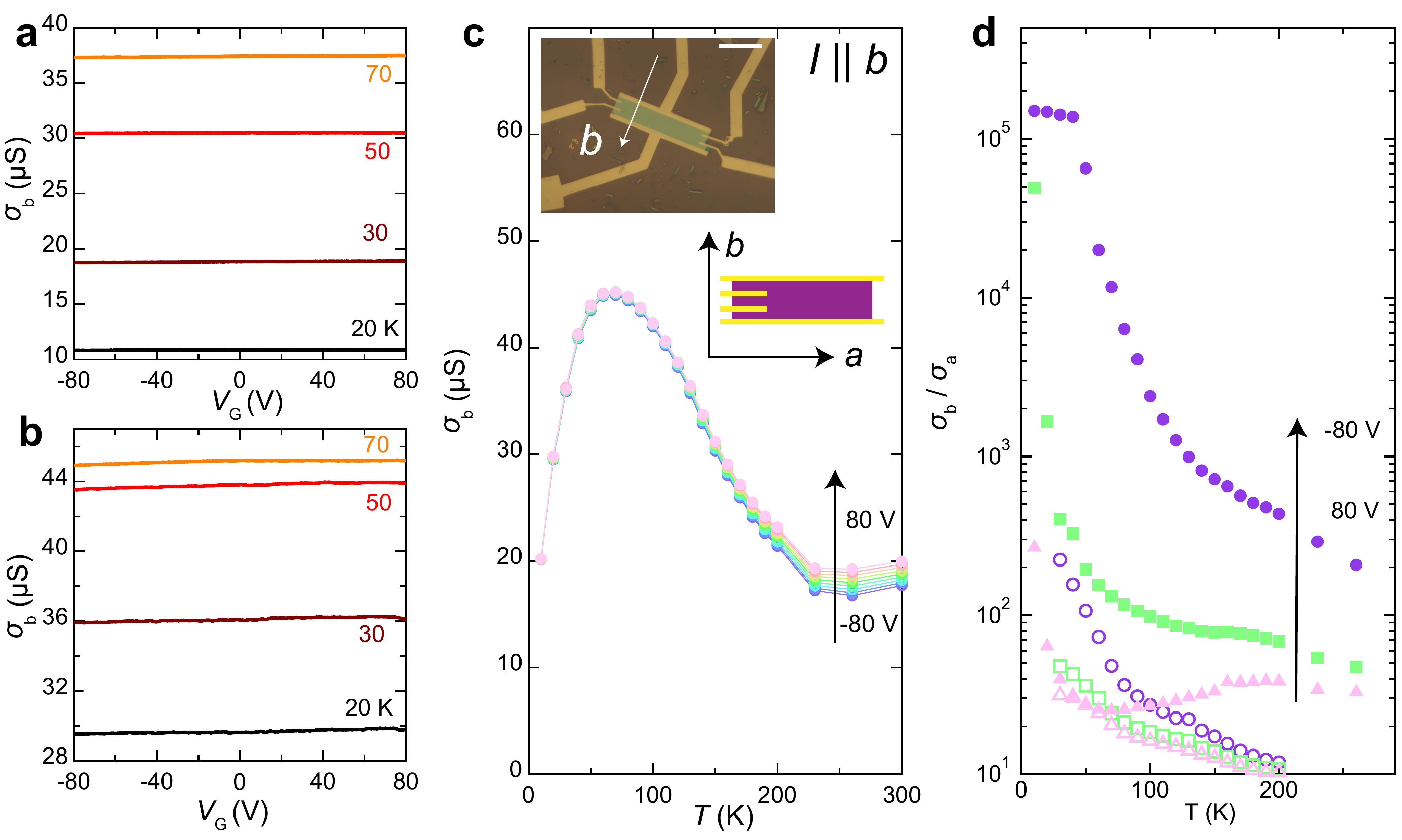}
	\caption{\textbf{Transport characteristics of CrSBr in the \textit{b} direction.} \textbf{a,b} Conductivity $\sigma_b$ as a function of backgate $ V_{G} $ of two CrSBr multilayers respectively \textbf{a}) 30 nm and \textbf{b}) 12 nm thick (data shown for $T$ = 20, 30, 50 and 70 K). \textbf{c)} Temperature dependent conductivity $\sigma_b$ for the 12~nm device, from -80 V to 80 V with 20 V step size. The insets show the optical microscope image of the 30 nm CrSBr device (the scale bar is 20 $\mu$m) and the schematics of the measurement configuration. \textbf{d)} Conductivity anisotropy $\sigma_b$/$\sigma_a$ as a function of temperature at various backgate voltages ( -80 V, purple circles; 0 V, green squares; 80 V, pink triangles). Filled symbols represent data obtained by taking $\sigma_a$ and $\sigma_b$ measured on CrSBr multilayers whose data are shown in Fig. 2\textbf{a} and Fig. 3\textbf{b}; the open symbols represent data obtained by taking $\sigma_a$ and $\sigma_b$ measured on the CrSBr multilayers whose data are shown in Fig. 2\textbf{b} and Fig. 3\textbf{b}. Note that for large negative gate voltage ( $V_G$ = -80 V) and at low temperature the anisotropy fluctuates in different devices, but it is always orders of magnitude larger than the conductivity anisotropy observed in any other 2D semiconductor: $10^5$ in the first case (full symbols) and $3\cdot10^2$ in the second case (open symbols).}
	\label{fig:03}
\end{figure*}

The devices were mounted in the vacuum chamber of a cryostat enabling transport to be studied as a function of temperature $T$, applied magnetic field $\mu_0 H$, and gate voltage \vg. In the experiments, we measured different devices to compare the $T$ and \vg\ dependence of the conductivity along the \textit{a}- and \textit{b}-direction, and we found a strikingly different qualitative behavior. The difference can be easily appreciated by comparing the measurements plotted in \textbf{Figure~\ref{fig:02}} (transport in the \textit{a}-direction) and \textbf{Figure~\ref{fig:03}} (transport in the \textit{b}-direction). In each figure we show data for two different devices that illustrate the common aspects of the observed behavior, as well as aspects that depend on the specific device.\\

We start discussing the transport in the \textit{a}-direction. The conductivity $\sigma_a$ measured as a function of \vg\ at different temperatures $T < 100$~K is shown in \textbf{Figure~\ref{fig:02}a} and \textbf{b}, for two devices realized on multilayers that are respectively 6 and 12~nm thick. In both cases $\sigma_a$ decreases systematically upon lowering $T$ and upon applying a negative \vg, a trend that is common to all devices measured. The absolute value of $\sigma_a$ at \vg~=~0~V and how strongly $\sigma_a$ is suppressed by the application of a negative $V_G$ depend on the device, with the differences likely originating from the different multilayer thicknesses (thicker devices are less sensitive to the applied gate voltage) and from material inhomogeneity (\ie\ inhomogeneity in the concentration of dopants unintentionally present in the material). Irrespective of these differences, at sufficiently large negative \vg\ the conductivity in the \textit{a}-direction is completely --or nearly completely-- suppressed in all cases. \textbf{Figure~\ref{fig:02}c }shows the full $T$-dependence of $\sigma_a$ measured at different \vg\ values between room temperature and $T= 10$~K, for the device whose low-$T$ data are shown in \textbf{Figure~\ref{fig:02}a}. At $V_G=0$~V, $\sigma_a$ is strongly reduced upon cooling from 300 to 10~K, another aspect common to all devices investigated. When $\sigma_a$ is plotted in logarithmic scale versus $1/T$, a linear relation is found, characteristic of a thermally activated Arrhenius behavior (\textbf{Figure~\ref{fig:02}d}). The activation energy extracted for $T<100$~K increases from 30 to approximately 300~meV upon applying an increasingly large negative \vg\ (\textbf{Figure~\ref{fig:03}e}).\\

The qualitative behavior exhibited by the low-temperature conductivity in the \textit{a}-direction may appear typical of what is expected from a doped semiconductor, and indeed it is in this way that earlier studies interpreted the experimental results~\cite{telford_layered_2020}. However, measurements of the conductivity $\sigma_b$ in the \textit{b}-direction indicate that the electronic properties of CrSBr are not at all consistent with such a scenario. $\sigma_b$ for two devices based on multilayers that are 30~nm and 12~nm thick is plotted in \textbf{Figure~\ref{fig:03}a} and \textbf{b}, as a function of \vg\ for several different temperatures below 100 K. The data show that the dependence of $\sigma_b$ on $T$ and \vg\ is strikingly different from that of $\sigma_a$. $\sigma_b$ depends only weakly on $T$, and is virtually independent of \vg: at $T=10$~K $\sigma_b \approx e^2/h$ for all \vg, comparable to the value measured at room temperature. This is even clearer in \textbf{Figure~\ref{fig:03}c}, which shows the full temperature dependence (from 300 to 10~K) measured for different values of \vg.\\

The absence of thermally activated transport in the \textit{b}-direction and the insensitivity of $\sigma_b$ to \vg\ implies that transport in the \textit{a}-direction (see \textbf{Figure~\ref{fig:02}}) cannot be interpreted in terms of temperature or gate-voltage induced depletion of charge carriers (\ie\ the common behavior of doped semiconductors). Indeed, if charge carriers were depleted by applying a gate voltage, or if they would be trapped at low temperature around the dopants where they originate from, a clear suppression of transport should be observed not only in $\sigma_a$, but also in $\sigma_b$. This is however not the case. Interestingly, as a result of the different $T$ and \vg\ dependence of $\sigma_a$ and $\sigma_b$, the ratio $\sigma_b/\sigma_a$, plotted in \textbf{Figure~\ref{fig:03}d}, becomes extremely large at low temperature and negative gate voltage, reaching values $\approx 10^5$ (the precise value ranges between $3 \cdot10^2$ and $10^5$ and depend on which device is used to measure $\sigma_a$ and $\sigma_b$, see open and filled symbols in \textbf{Figure~\ref{fig:03}d}). Such a large $\sigma_b/\sigma_a$ ratio is approximately 2 to 5 orders of magnitude larger than the anisotropy in low-frequency in-plane conductivity measured in any other 2D van der Waals semiconductor, as recently reviewed in Ref. \cite{zhao_-plane_2020}. We conclude that both qualitative and quantitatively $T$- and \vg-dependent conductivity measurements exhibit an extremely anisotropic behavior, and appear incompatible with conventional scenarios based on band transport in a doped 2D semiconductor.\\

Albeit an anisotropy is clearly visible in \textit{ab-initio} calculation of the band structure (see \textbf{Figure~\ref{fig:01}b}), such a large difference in the conductivity along the \textit{a}- and \textit{b}-direction is unexpected when looking at the material structure, which does not appear to be particularly anisotropic. That the relation between electronic anisotropy and structure is unusual is also shown by the observation that the long axis of exfoliated CrSBr multilayers (which are platelets having one of their sides much longer than the other) corresponds to the crystallographic \textit{a}-direction, \ie\ to the direction with low electrical conductivity. This is indeed rather unique, as it is commonly the case that materials exhibiting strongly anisotropic electronic transport are needles whose long axis points in the high conductivity direction (see, for examples, conductors such as ReS$_2$, NbSe$_3$, TiS$_3$, ZrTe$_5$, black phosphorus or high-mobility organic semiconductors such as rubrene~\cite{kurita_current_2000,sundar_elastomeric_2004,reese_organic_2007,liu_integrated_2015,tao_mechanical_2015,qiu_observation_2016,hong_polarized_2014,khatibi_anisotropic_2020,patra_anisotropic_2020}).\\

\begin{figure*}
	\centering
	\includegraphics[width=0.9\linewidth]{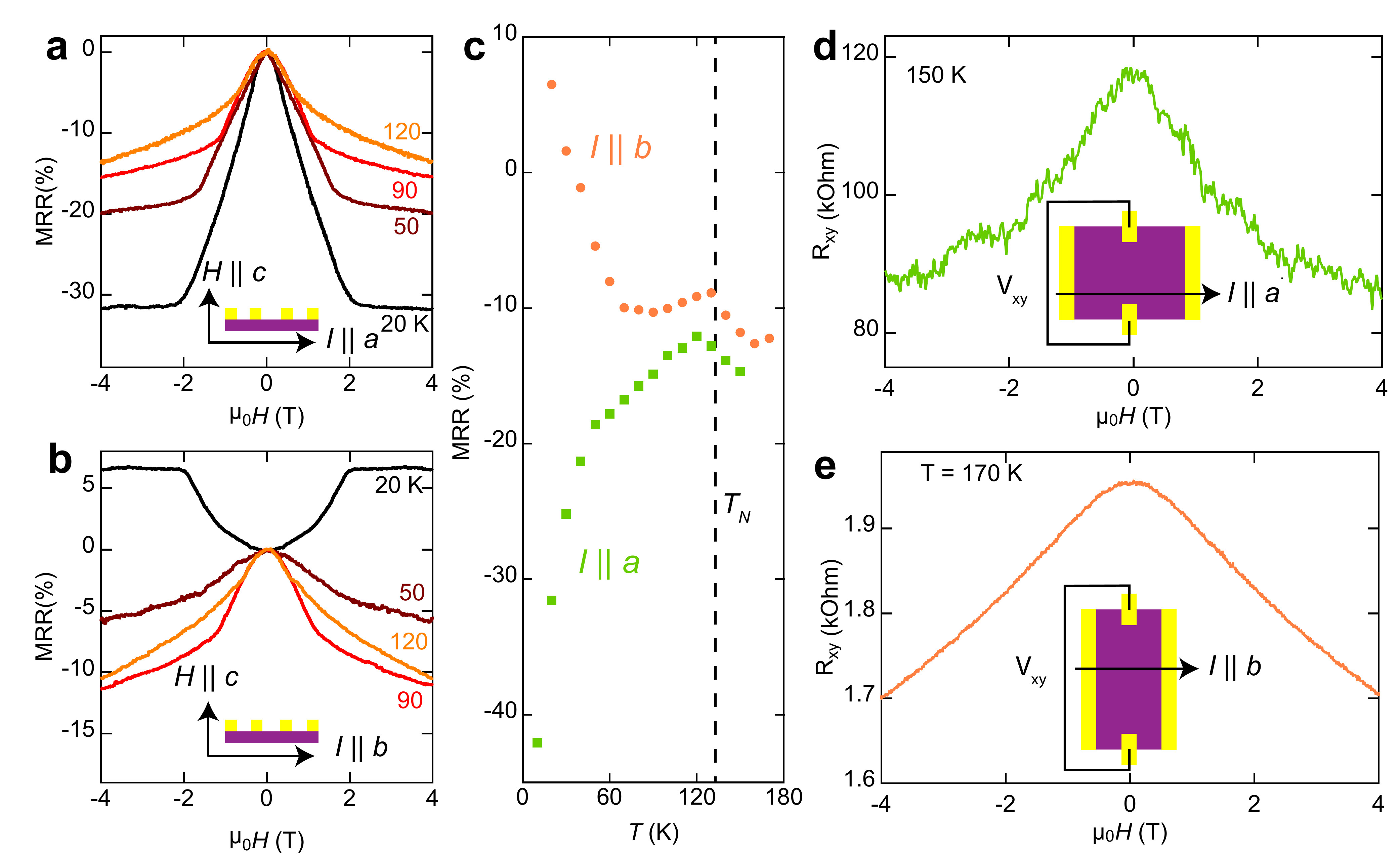}
	\caption{\textbf{Magnetotransport properties of CrSBr multilayers.} \textbf{a,b} Magnetoresistance ratio (MRR $= (R(\mu_0H)-R(0))/R(0)$) measured at different temperatures below the \emph{Néel}-temperature ($T_N = 132$ K), as indicated in the legend, respectively along the \textit{a}- and \textit{b}-direction (data taken on the devices whose \vg\ dependent transport characteristics are shown in Fig. 2\textbf{a} and Fig. 3\textbf{a}. Note the change in sign of the magnetoreistance that is observed in the 20~K curve when measuring in the \textit{b}-direction (\ie\ panel \textbf{b}). The insets show the measurement configuration and the direction of the applied magnetic field. \textbf{c)} Magnetoresistance raito MRR measured in the \textit{a}- and \textit{b}-directions at $\mu_0 H = 4$ T. \textbf{d,e)} Transverse magnetoresistance measured with current flowing along the \textit{a}- and \textit{b}-direction, respectively (data taken at 150~K and 170~K; the magnetic field is applied in the \textit{c} direction). No sign of a component of the transverse resistance antisymmetric in the applied field is present in the data. The schemes in the insets illustrate the measurement configurations.}
	\label{fig:04}
\end{figure*}

We found further indications of the unusual transport behavior of CrSBr multilayers in magnetotransport measurements performed with the magnetic field applied perpendicular to the CrSBr layers. Below the \emph{Néel} transition temperature ($T_N= 132$~K), the longitudinal magnetoconductivity in the \textit{a}- and \textit{b}-direction exhibits a qualitatively different temperature dependence. In the \textit{a}-direction (see \textbf{Figure~\ref{fig:04}a}), the magneoresistance is negative (\ie\ the resistance decreases upon the application of an external magnetic field) and increases in magnitude as temperature is lowered, showing a dependence on $T$ and $\mu_0 H$ that conforms to results reported in earlier studies~\cite{telford_layered_2020,telford_hidden_2021}. In the \textit{b}-direction (see \textbf{Figure~\ref{fig:04}b}), instead, the magnetoresistance is initially negative (for $T<T_N$), it decreases in magnitude, and eventually inverts its sign at approximately 30~K (possibly in concomitance with a additional magnetic transition as also detected by bulk magnetization measurements~\cite{telford_layered_2020,telford_hidden_2021}; see \textbf{Figure S4}, Supplementary Information). At even lower temperature the magnetoresistance is positive and keeps increasing in magnitude upon further cooling. The overall temperature dependence of the magnetoresistance in the \textit{a}- and \textit{b}-direction --shown in \textbf{Figure~\ref{fig:04}c}-- is an additional manifestation of the qualitatively different nature of transport in the two directions (note for completeness that, irrespective of temperature, no significant hysteresis in the magnetoresistance was observed upon sweeping the magnetic field up and down, neither in the \textit{a}- nor in the \textit{b}- direction).\\

A possibly even more striking outcome of magnetotransport measurements is the virtually complete absence of a component of the transverse resistance antisymmetric with respect to the applied magnetic field, \ie\ the absence of Hall effect. Irrespective of whether the current is sent in the \textit{a}- or \textit{b}-direction, the transverse voltage that we measure is symmetric in the applied magnetic field (see \textbf{Figure~\ref{fig:04}d} and \textbf{e}), and likely originates from the Hall probes picking up part of the longitudinal magnetoresistance. After antisymmetrization, the transverse resistance results in a vanishingly small signal comparable to (or smaller than) the noise of the measurements, irrespective of the temperature and of the gate voltage at which measurements are performed (see detailed discussion section S3 in Supplementary Information){gate voltage at which measurements are performed and of the temperature, \ie\ the Hall effect is absent regardless of whether CrSBr is in its layered antiferromagnetic state for $T<T_N$ or in its paramagnetic state for $T>T_N$ (see also Section S3 in the Supplementary Information, and \textbf{Figure~S8} in there for more details)}. In line with all other observations presented above, the absence of Hall signal confirms that conduction in CrSBr is not due to coherent 2D band transport, as it may be expected for a conventional doped 2D semiconductor.\\

Having established that low-frequency transport exhibits a behavior inconsistent with conventional band conduction of a doped 2D semiconductors, we discuss photocurrent measurements that provide a clear indication as to the physical origin of the unusual observed behavior. We measure photocurrent in the absence of an externally applied bias, by illuminating the device with a tunable laser source that excites electron-hole pairs (possibly bound into excitons, although this is not a crucial aspect here, because in thick multilayers electrostatic screening strongly suppresses excitonic effects \cite{wang_colloquium_2018,zultak_ultra-thin_2020,carre_excitons_2021}). In the experiments, the laser is focused on a spot of approximately 2~\mum\ diameter and, as it is commonly the case~\cite{ahn_scanning_2005,ubrig_scanning_2014}, a finite photocurrent is measured only when the laser spot is positioned sufficiently close to one of the metal contacts. The photocurrent is due to minority photo-excited carriers that escape from the sample into the nearby contact, and --if the laser spot is located away from the contacts-- the photocurrent vanishes because electron-hole pairs recombine before the carriers have the time to escape~\cite{ahn_scanning_2005}. In this regime, the photocurrent probes light absorption by interband transitions, as discussed more in detail in Section S5 of the Supplementary Information. It is therefore proportional to the so-called joint density of states of the two bands involved in the transition, at the energy of the incident photons~\cite{miller_electric_1985,collins_photocurrent_1986}. In other words, measuring the photocurrent as a function of wavelength of incident photons provides information analogous to what can be inferred from a wavelength-dependent optical absorption measurements.\\

\begin{figure}
	\includegraphics[width=.99\linewidth]{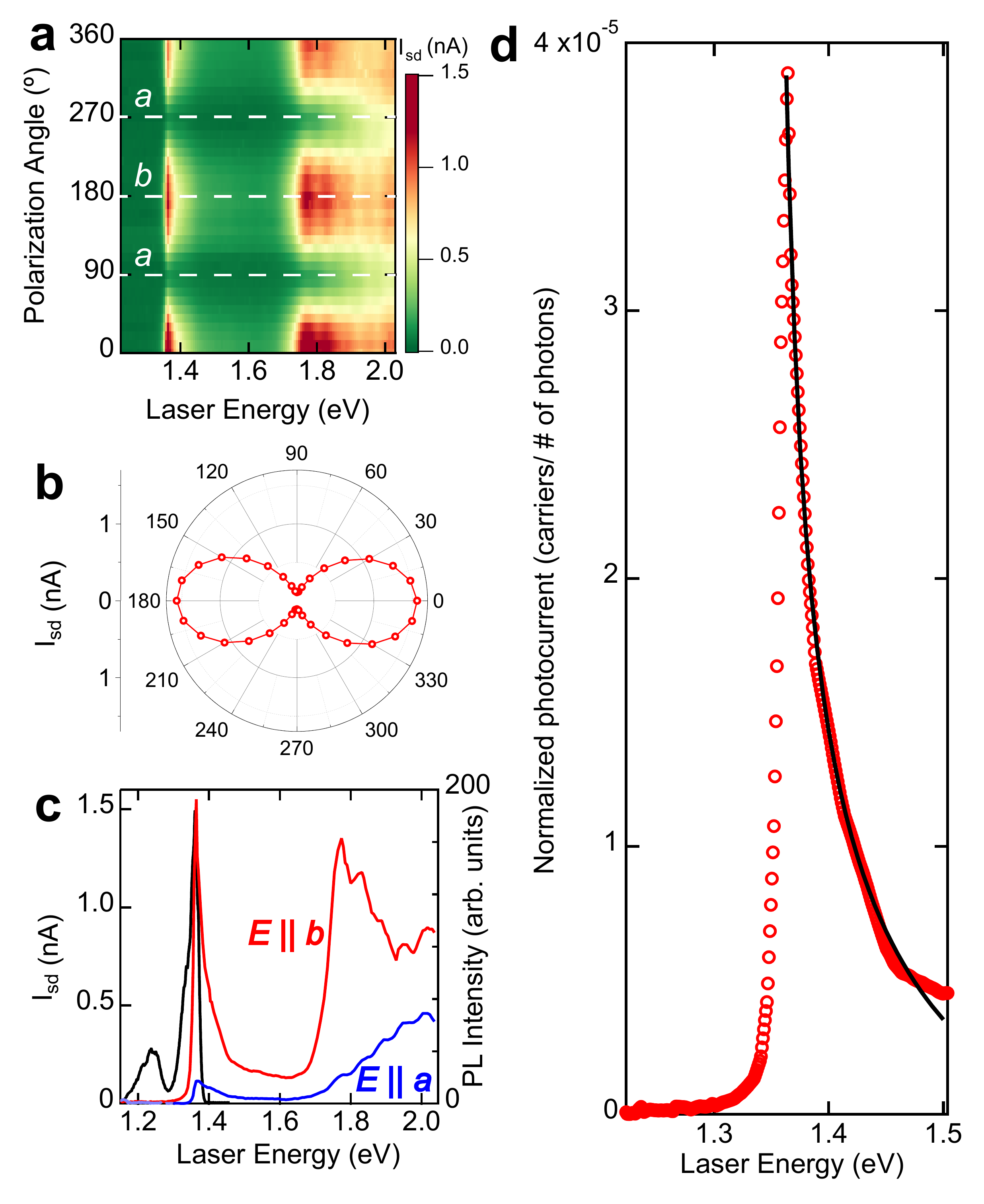}
	\centering
	\caption{\textbf{Photocurrent measurements on CrSBr.} \textbf{a)} Color plot of the photocurrent (measured in the absence of applied bias at T~=~10~K) as function of laser energy and polarization angle relative to the \textit{b} crystallographic direction of CrSBr; the white dashed lines correspond to the light polarization pointing along the \textit{a}- and \textit{b}-direction, as noted on the plot. \textbf{b)} Polar plot of the photocurrent (measured under illumination with linearly polarized light at $\lambda$ = 910 nm), as a function of angle relative to the \textit{b} crystallographic direction of CrSBr. \textbf{c)} Photocurrent versus photon energy for light polarization pointing along the \textit{a}- (blue) and \textit{b}- (red) directions (left vertical axis), showing sharp peaks due to inter-band transitions at 1.35 and 1.75~eV; the black line is the photoluminescence spectrum (right vertical axis) for incident light polarized along the \textit{b}-direction. \textbf{d)} Normalized photocurrent as a function of photon energy, measured with light polarized along the \textit{b}-direction (red open circles). To properly analyze the spectral dependence in terms of joint density of states,  the photocurrent measured at any given wavelength is normalized to the rate of incident photons at that wavelength (the photocurrent normalized in this way corresponds to the the so-called external quantum efficiency).} The black line is a fit with a $ 1/\sqrt{E-E^*}$ functional dependence ($E^*$ = 1.35 eV), as expected for a \emph{van Hove} singularity of a 1D quadratically dispersing band.
	\label{fig:05}
\end{figure}

The photocurrent measured as a function of photon energy and of angle between the light polarization vector and the \textit{b}-direction of the CrSBr crystal is shown in \textbf{Figure~\ref{fig:05}a}. Its intensity is maximum when the polarization is parallel to the \textit{b}-direction, and nearly vanishes when the polarization points in the \textit{a}-direction (most of the remnant signal likely originates from polarization leakage in our optical system). The photocurrent as a function of polarization angle is shown in the polar plot of \textbf{Figure~\ref{fig:05}b} (at fixed energy of the incident photon), which directly illustrates the presence of a very pronounced anisotropy. The anisotropy is so strong to suggest a quasi 1D electronic behavior, since finding that a sizable photocurrent is observed (i.e., that light is absorbed) only when the electric field of the incoming light points along the \textit{b}-direction can be understood if CrSBr behaves as a collection of decoupled 1D chains. Indeed, for decoupled 1D chains electrons are only able to absorb energy when they are accelerated in the direction parallel to the chain.\\

In agreement with this idea, we find that the main photocurrent peak observed at approximately $E^*=1.37$~eV (red line in \textbf{Figure~\ref{fig:05}c}, corresponding to the energy of the light measured in photoluminescence experiments; black line in \textbf{Figure~\ref{fig:05}c}) has a very pronounced asymmetry (red dots in \textbf{Figure~\ref{fig:05}d}) and its line shape past threshold is reproduced excellently by a $1/\sqrt{E-E^*}$ dependence (black curve in \textbf{Figure~\ref{fig:05}d}, measured at $T=10$ K). Such a functional dependence is the characteristic hallmark of the \emph{van Hove} singularity of a 1D electronic band \cite{marder_condensed_2010}, whose observation provides evidence of the one dimensional character of the electronic properties (more details about photocurrent and its relation to the joint density of states between 1D states can be found in Section S5 and S6 of the Supplementary Information). It is this one-dimensionality that is responsible for the unusual behavior of transport that we are reporting here. Without having performed a fully systematic study as a function of temperature, we find that the photocurrent signal increases rapidly in magnitude upon cooling the device from room temperature to $T=10$~K, with no change in its qualitative behavior above and below $T_N$. Consistently with all other transport anomalies of CrSBr discussed above --which are also present both for $T>T_N$ and for $T<T_N$-- this behavior indicates that the quasi 1D behavior of CrSBr does not originate from the transition into the magnetic state. \\

An extremely strong band structure anisotropy in CrSBr had been predicted by first-principles calculations, but the implication of such calculations for the interpretation of experiments can not be assessed on a purely theoretical basis. In particular, a strong anisotropy in the band structure does not give a clear indication as to the physical regime that needs to be considered to properly model and interpret transport measurements. On the contrary, finding that light is absorbed only when the electric field of the incoming wave points along the \textit{b}-direction, as well as the observation of a \emph{van Hove} singularity in the joint density of states, consistently indicate that it is more appropriate to rationalize experimental observations by viewing CrSBr as a quasi 1D electronic system, than as a strongly anisotropic 2D semiconductor. Indeed, even though much more work is needed to reach a detailed understanding of CrSBr, many of our experimental observations support a scenario in which CrSBr is described in terms of incoherently coupled 1D electronic chains, as we now proceed to discuss.\\

A scenario based on incoherently coupled 1D chains can rationalize why transport in the \textit{a}- and the \textit{b}-direction exhibits striking quantitative and qualitative differences. Indeed, measuring transport in the \textit{b}-direction corresponds to probing conduction along the chain, whereas transport in the \textit{a}-direction measures tunneling in between chains, and it is theoretically well-established that the two processes are sensitive to different physical phenomena~\cite{giamarchi_quantum_2003}. The appropriate conceptual framework to describe transport along the chain is the physics of Anderson localization, based on concepts such as phase coherence and interference~\cite{abrahams_50_2010}. A conductivity $\sigma_b \approx e^2/h$ implies that the system is still far from the fully localized state and that it is very sensitive to changes in the phase coherence time, which may be the reason for the unusual increase in conductivity seen upon lowering temperature from 200~K to 100~K (see \textbf{Figure~\ref{fig:03}c}; as the \emph{Néel}-temperature is approached, the diverging spin fluctuations enhance decoherence, thereby reducing the tendency to localize, causing a corresponding increase in conductivity). As compared to $\sigma_b$, the conductivity in the \textit{a}-direction is much more strongly suppressed as $T$ is lowered, because at low energy the tunneling probability is progressively reduced by electron-electron interactions~\cite{giamarchi_quantum_2003}, which accounts for the extremely large ratio $\sigma_b/\sigma_a$ observed at low $T$. Thinking of CrSBr in terms of incoherently coupled 1D chains also naturally explains the giant ratio $\sigma_b/\sigma_a$, the absence of Hall effect, and the \emph{van Hove} singularity observed in the photocurrent measurements. Owing to the different transport processes involved for motion along the \textit{a} and \textit{b} crystallographic directions, a model based on incoherently coupled 1D chains makes it possible to imagine scenarios to explain the different signs of the low-temperature magnetoconductivity in the two directions (simply because inter-chain tunneling and quantum interference along the chain respond differently to an applied magnetic field).\\

What remains to be identified is a mechanism explaining why transport in the \textit{a}- and \textit{b}-direction exhibit such a different dependence on gate voltage. Such a mesmerizing behavior was never reported earlier for any other semiconducting material, which currently prevents its explanation by analogy with any other known electronic system. Nevertheless --as it appears impossible to explain the different gate voltage dependence of the conductivities $\sigma_a$ and $\sigma_b$ in terms of conventional band transport of a 2D doped semiconductor-- this observation also points to the need to invoke a different microscopic scenario to understand the nature of electronic motion on CrSBr. Our claim that transport should be described in terms of a collection of incoherently coupled 1D chains is consistent with this conclusion.\\

\section{Conclusion}
CrSBr has been recently identified as an extremely interesting 2D magnetic semiconductor, rather unique because the width of its bands is approximately 1.5~eV or larger, \ie\ 50-100 times larger than that of other semiconducting 2D magnetic materials that have been recently studied~\cite{wang_electronic_2011,wilson_interlayer_2021,yang_triaxial_2021}. Earlier work has mainly focused on the magnetic properties of the material, exploiting the fact that the large bandwidth results in a conductivity that is also sufficiently large to remain measurable at low temperature, well below the magnetic transition temperature. Our work concentrates on the electrical transport properties of the material that have so far received less attention, and shows that they also exhibit strikingly unique aspects, resulting from an extremely strong anisotropy. Such a strong anisotropy manifests itself in a qualitatively different gate-voltage, temperature, and magnetic field dependence of the conductivity along the \textit{a}- and \textit{b}- directions, in the complete absence of Hall effect for all temperatures and gate voltages investigated, as well as in the (nearly complete) absence of photocurrent under illumination with light polarized in the \textit{a} crystallographic direction. These observations, as well as the giant anisotropy $\sigma_b/\sigma_a > 3\cdot10^2$ --orders of magnitude stronger than in any known semiconductor-- are consistent with the strong anisotropy found in \textit{ab-initio} calculations of the band structure, but incompatible with a description of the observed phenomena in terms of conventional band transport of a doped semiconductor. In contrast, all the phenomenology found in our experiments --including the manifestation of a \emph{van Hove} singularity characteristic for 1D quadratically dispersing bands-- appears to be much more naturally interpreted by viewing the electronic properties of CrSBr as originating from the presence of weakly and incoherently coupled 1D chains. Even though many detailed aspects of conduction in the material remain to be understood--and most notably the very different gate voltage dependence of the conductivity in the \textit{a}- and \textit{b}-directions-- we can conclude based on our experiments that CrSBr represents a first example of 2D van der Waals magnetic semiconductor, in which electrical transport occurs in the quasi 1D regime.\\

\medskip
\textbf{Supporting Information} \par 
Supporting Information is available from the Wiley Online Library or from the author.

\medskip
\textbf{Acknowledgements} \par 
We gratefully acknowledge Alexandre Ferreira for continuous and precious technical support, Catherine Witteveen and Vanessa Kronenberg for experimental help during the synthesis of this material and Enrico Giannini, Tierry Giamarchi for fruitful discussion. AFM gratefully acknowledges financial support from the Swiss National Science Foundation (Division II) and from the EU Graphene Flagship project. FVR acknowledges Swiss National Science Foundation under Grant No. PCEFP2 194183. MG acknowledges support from the Italian Ministry for University and Research through the Levi-Montalcini program. KW and TT acknowledge support from the Elemental Strategy Initiative conducted by the MEXT, Japan (Grant Number JPMXP0112101001) and JSPS KAKENHI (Grant Numbers 19H05790, 20H00354 and 21H05233).

\medskip

%

%



\end{document}